\documentclass[twocolumn,nofootinbib,showpacs,preprintnumbers,amsmath,amssymb,epsfig,widetext]{revtex4}

\usepackage{graphicx}
\usepackage{dcolumn}
\usepackage{bm}
\usepackage{epsfig}
\usepackage{natbib}

\def\la{\mathrel{\mathpalette\fun <}}
\def\ga{\mathrel{\mathpalette\fun >}}
\def\fun#1#2{\lower3.6pt\vbox{\baselineskip0pt\lineskip.9pt
  \ialign{$\mathsurround=0pt#1\hfil##\hfil$\crcr#2\crcr\sim\crcr}}}

\input epsf

\newcommand {\boldtheta}{\mbox{\boldmath $\theta$}}
\def\be{\begin{equation}}
\def\ee{\end{equation}}
\def\ba{\begin{eqnarray}}
\def\ea{\end{eqnarray}}

\newcommand{\mnras}{Mon. Not. R. Astron. Soc.}

\begin{document}

\preprint{}

\title{Cosmological Constraints on $f(R)$ Acceleration Models}

\author{Yong-Seon Song, Hiranya Peiris,\footnote{Hubble Fellow} and Wayne Hu}
\email{ysong@kicp.uchicago.edu} 
\affiliation{Kavli Institute for Cosmological Physics, Department of Astronomy \& Astrophysics, and Enrico Fermi Institute,  University of Chicago, Chicago IL 60637 \\
}

\date{\today}

\begin{abstract}
Models which accelerate the expansion of the universe through the addition of a function of the Ricci scalar $f(R)$
leave a characteristic signature in the large-scale structure of the universe at the Compton wavelength
scale of the extra scalar degree of freedom.  We search for such a signature in current cosmological
data sets: the WMAP cosmic microwave background (CMB) power spectrum, SNLS supernovae distance
measures, the SDSS luminous red galaxy power spectrum, and galaxy-CMB angular correlations.   
Due to theoretical uncertainties in the nonlinear evolution of $f(R)$ models, the galaxy power spectrum
conservatively
yields only weak constraints on the models despite the strong predicted signature
in the linear matter power spectrum.   Currently the tightest constraints involve the modification to the
integrated Sachs-Wolfe effect from growth of gravitational potentials during the acceleration epoch.
This effect is manifest for large Compton wavelengths 
in enhanced low multipole power in the CMB and anti-correlation between 
the CMB and tracers of the potential.  They place a bound on the
 Compton wavelength of the field be less than of order the Hubble scale.
\end{abstract}

\maketitle

\section{introduction} \label{sec:intro}

Cosmic acceleration can arise either from an unknown component of dark energy with
negative pressure or a modification to gravity that only appears at cosmological
scales and densities.   Additional terms in the Einstein-Hilbert action 
that are non-linear functions $f(R)$ of the Ricci scalar $R$ have long been known
to cause acceleration \cite{Sta80,Caretal03,CapCarTro03,NojOdi03} and have been
the subject of much recent interest 
\cite{Poplawski:2006kv,delaCruz-Dombriz:2006fj,Brookfield:2006mq,Sotiriou:2006sf,Sotiriou:2006hs,Bean:2006up,Amendola:2006we,Baghram:2007df,Bazeia:2007jj,Li:2007xn,Bludman:2007kg,Rador:2007wq,Faraoni:2006sy,Koivisto:2006ie,Capozziello:2007gm,Nojiri:2006gh,Nojiri:2006su,Capozziello:2006dj,Fay:2007uy,Amendola:2007nt,SonHuSaw06,SawHu07,Tsujikawa:2007gd,Appleby:2007vb,Charmousis:2007ji,DeFelice:2007ez}.

The main challenge for $f(R)$ models as a complete description of gravity 
lies with the extremely tight constraints on such modifications placed by 
solar system and local tests of general relativity.  Chiba \cite{Chi03} showed that the
fundamental problem is that $f(R)$ models generically introduce a light scalar degree of
freedom with a long Compton wavelength at cosmological densities \cite{Chiba:2006jp, EriSmiKam06}.    This problem can be mitigated by the so-called chameleon 
mechanism \cite{KhoWel04,MotBar04}, where the local Compton wavelength can decrease in high
density environments \cite{NavAco06,FauTegBun06}.  
The cosmological Compton wavelength can then be a scale of cosmological interest.  
Nevertheless, it must typically be less than
a few tens of Mpc if the Galactic gravitational potential 
is not substantially deeper than implied by local rotation curve measurements \cite{HuSaw07}.

In this {\it Paper}, we take the perspective that $f(R)$ models are effective theories
that are valid at cosmological densities and scales and are not necessarily predictive
at the high densities and small scales of local tests.  Hence we explore models
with Compton wavelengths out to the horizon size and seek to constrain them from
cosmological
observables alone. 
At the very least, this exploration yields a cosmological test of general relativity that is
independent of local constraints.

Although $f(R)$ models can change the expansion history during the
acceleration epoch, there in principle always exists a dark energy model that provides
the same history.
The unique and strongest signatures of $f(R)$ modifications are on cosmological structure formation \cite{SonHuSaw06} and can impact observables down to Compton wavelengths of a 
few Mpc \cite{HuSaw07}.
We consider constraints arising from 
the CMB angular power spectrum measured by WMAP \cite{Speetal06}, the linear matter power spectrum inferred from the Sloan Digital Sky Survey (SDSS) luminous red galaxy (LRG) sample \cite{Tegetal06}, the distance measures by  the Supernovae Legacy Survey (SNLS) \cite{Astier}, and the cross-correlation between the CMB and large scale structure
 as measured by WMAP and a range of galaxy and quasar surveys \cite{AfsLohStr04, BouCri04, FosGaz04, Giannantonio06, Cabre06}.

The outline of the paper is as follows.  In \S \ref{sec:methodology}, we review the 
$f(R)$ model and discuss our calculation and analysis methods.  In \S \ref{sec:constraints}
we present the results of joint cosmological constraints on these models.  We discuss
these results in \S \ref{sec:discussion}.

\section{$f(R)$ Methodology}
\label{sec:methodology}

\subsection{Model}
\label{sec:parameters}

In $f(R)$ models of gravity, the Einstein-Hilbert action is supplemented by
a term that is non-linear in the Ricci scalar $R$
\ba
S=\int d^4x \sqrt{-g}\left[ {R+f(R) \over16\pi G}+{\cal L}_{\rm m} \right]\,,
\ea
where ${\cal L}_m$ is the matter Lagrangian.  The modified Einstein and
 Friedmann equations
result from varying the action with respect to the metric.
Given the freedom to choose a functional form for $f(R)$, any desired
expansion history can be replicated
\cite{Multamaki:2005zs,CapNojOdiTro06,Nojiri:2006gh,NojOdi06,delaCruz-Dombriz:2006fj,SonHuSaw06}.
In particular, one can choose the $\Lambda$CDM expansion history which is
known to satisfy distance constraints from high redshift supernovae, 
baryon acoustic oscillations and the CMB. 

Even given the degeneracy with dark energy in the expansion history, $f(R)$ 
models have  distinguishable effects on the formation of structure.
The promotion of the Ricci scalar to a dynamical degree of freedom modifies
the force law between particles.  This modified force is mediated by a
new scalar degree of freedom $f_{R} \equiv d f/d R$, which has a squared
Compton wavelength proportional to  $f_{RR} \equiv d^{2}f/dR^{2}$.  
Below the Compton wavelength scale gravity becomes a scalar-tensor
theory, leading to an enhancement in the growth of cosmological
density perturbations and a corresponding suppression in the decay
of gravitational potentials.

For cosmological tests, it is convenient to express the squared
Compton wavelength in the background in units of the Hubble length 
squared \cite{Zha05,SonHuSaw06}
\begin{eqnarray}
B \equiv {f_{RR} \over 1+f_R} {d R \over d\ln a} \left(  d\ln H \over d\ln a \right)^{-1}\,,
\end{eqnarray}
where $a$ is the scale factor and the Ricci scalar is evaluated at the background density.
We will specialize our consideration to $f(R)$ models that exactly 
reproduce the $\Lambda$CDM expansion history to test whether
the unique signatures of $f(R)$ gravity are seen in current 
cosmological data sets.  

Under this assumption, the additional
degree of freedom in $f(R)$ gravity is parameterized by the value of $B$ today,
$B_{0} \equiv B(z=0)$.  
Stability requires the mass squared of the
scalar to be positive ({\it i.e.}~a prior of $B_{0}\ge 0$) \cite{SonHuSaw06,SawHu07}.
Note that in the limit that $B_{0} \rightarrow 0$, the phenomenology of
$\Lambda$CDM is recovered
in structure formation tests as well as expansion history tests.
More generally, the control parameter is the average Compton wavelength
through the acceleration epoch when gravity is modified.

\subsection{Power Spectra Calculation}

 The fundamental observables of our $f(R)$ model are the same as in $\Lambda$CDM.
 These include the redshift-distance
 relation, the CMB angular power spectrum $C_\ell$, galaxy power spectra 
 $P_{g}(k)$, and the angular correlation between galaxies and the CMB $w(\theta)$.  
 Cosmic shear, dark matter halo profiles and masses from weak lensing as well as
  the cluster abundance
 are also potential observables
 but their utilization requires cosmological simulations that are beyond the scope of this
 work.
 
We modified the Boltzmann code CAMB \cite{Lewetal00} to calculate these observables
 in  $f(R)$ gravity.  
In the CAMB code, the density perturbations are computed in the synchronous gauge. We evolve the 
usual Boltzmann code up to $a\sim 0.01$ when deviations introduced by $f(R)$ are
still small. At this epoch, we transform the matter perturbations from synchronous gauge to gauge invariant variables, and feed them as initial conditions to 
the linear perturbation equations for the density fluctuation and CMB sources (see
\cite{SonHuSaw06} Eqs. 28-35).
Once the initial conditions are specified, the power spectrum observables are computed from a separate $f(R)$ routine that bypasses the usual CAMB code 
without significantly increasing the computational time.   

As we shall discuss, the main modification made by our $f(R)$ models on the CMB is a change
in the evolution of gravitational potentials during the acceleration epoch.   The CMB power
spectrum is modified at low multipoles through the so called ``Integrated Sachs-Wolfe" (ISW) 
effect, and the temperature field is correlated with tracers of gravitational potentials such
as galaxies.  This correlation is strong if the redshift of the galaxies is matched to the epoch
at which the gravitational potentials evolve.
Following \cite{LoVerde:2006cj}, we model the angular correlation between a set of galaxy
surveys indexed by $i$ and the CMB,
assuming a broad band selection function for the galaxies,
\ba
n_{i}(z)=\frac{1.5}{\Gamma[2]}\frac{z^2}{z_i^3}e^{-(z/z_i)^{1.5}}\,,
\ea
where the $z_i$ are chosen to match the median redshift in the surveys.
While crude, the uncertainties introduced by this choice of selection function are smaller
than the difference in model predictions that will be considered.
The galaxy bias is assumed to be constant in each redshift bin.  We calculate
the cross power spectrum between galaxy number density and CMB temperature
following \cite{SonHuSaw06} (Eqs. 55-60) and transform
it to the angular power correlation function $w_{i}(\theta)$.

\subsection{Likelihood Analysis}

We use a Markov Chain Monte Carlo (MCMC) technique \cite{Christensen:2000ji,
Christensen:2001gj,Knox:2001fz,LewBri02,Kosowsky:2002zt,Verde:2003ey} to
evaluate the likelihood function of model parameters. The MCMC is used to
simulate observations from the posterior distribution ${\cal P}(
\boldtheta|x)$, of a set of parameters $\boldtheta$ given event $x$, obtained
via Bayes' Theorem,
\begin{equation}
{\cal P}(\mbox{\boldmath $\boldtheta$}|x)=
\frac{{\cal P}(x|\boldtheta){\cal P}(\boldtheta)}{\int
{\cal P}(x|\boldtheta){\cal P}(\boldtheta)d\boldtheta},
\label{eq:bayes}
\end{equation}
\noindent where ${\cal P}(x|\boldtheta)$ is the likelihood of
event $x$ given the model parameters $\boldtheta$ and ${\cal
P}(\boldtheta)$ is the prior probability density. The MCMC
generates random draws (i.e.~simulations) from the posterior distribution that
are a ``fair'' sample of the likelihood surface. From this sample, we can
estimate all of the quantities of interest about the posterior distribution
(mean, variance, confidence levels). A properly derived and implemented MCMC
draws from the joint posterior density ${\cal P}(\boldtheta|x)$
once it has converged to the stationary distribution. We use 16 chains and a conservative Gelman-Rubin convergence criterion \cite{gelman/rubin} to determine when the chains have converged to the stationary distribution.

For our application, $\boldtheta$ denotes a set of cosmological parameters. We then use a modified version of the CosmoMC code \cite{LewBri02} to determine constraints placed on this parameter space by WMAP, SDSS LRG galaxy power spectrum, and supernova distance measures. For the LRG data, we only use the first 14 $k$-bins in our fits reflecting a conservative cut for linearity $k < 0.1 h$/Mpc. 
For the supernovae constraint, we take the Supernovae Legacy Survey data set and its analysis remains unaffected by our $f(R)$ modification.  Likewise, its impact is mainly to help determine expansion history parameters.
In our analysis, we take the parameter set  $\{\omega_b\equiv \Omega_b h^2$, $\omega_m
\equiv \Omega_m h^2$, $\theta_A, \ln(10^{10}A_{s}), n_s, \tau, B_0, b, Q_{\rm nl}\}$, where $\theta_A$ is the angular size of the acoustic horizon, and $A_{s}$ is the power in the primordial curvature perturbation at $k=0.05\ {\rm Mpc}^{-1}$. The universe is assumed to be to be spatially flat.  Recall also that we
fix the expansion history to be the same as a $\Lambda$CDM model.  Hence $\theta_{A}$ is also
a proxy for the Hubble constant $H_0\equiv 100h$ km/s/Mpc or $\Omega_{m}$.

The last two parameters in the set are the linear galaxy bias $b$ and non-linearity parameter $Q_{\rm nl}$ required for the interpretation of the galaxy power spectrum $P_g(k)$ of SDSS LRGs \cite{Tegetal06} (see also \S \ref{sec:Pk}), defined as
\begin{equation}
P_g(k) = P_{L}(k)\ b^2\frac{1+Q_{\rm nl} k^2}{1+1.4 k}.
\label{eq:b_Qnl}
\end{equation}
The linear bias factor $b$ is defined with respect to the linear matter power spectrum $P_{L}(k)$
at $z=0.35$, the effective 
redshift of the LRG galaxies. Since $f(R)$ models predict a scale dependent 
growth rate that changes the shape of the power spectrum as a function of redshift, 
we do not define the bias as relative to the power spectrum at $z=0$ ({\it cf.}~\cite{Tegetal06}). 
The second factor on the right hand side accounts for the non-linear evolution of the matter power spectrum shape and the scale-dependent bias of the galaxies relative to dark matter.  We assume that this \emph{ansatz} continues to be a valid approximation for $f(R)$ theories. We do not marginalize analytically over the $b$ and $Q_{\rm nl}$ parameters as is done normally in the CosmoMC code; instead the parameters are marginalized numerically as independent nuisance parameters in the MCMC analysis. 

For all of our parameters except for $B_0$, we employ flat linear priors
in the stated parameters.  For $B_0$, we supplement
the flat prior with the stability condition that  $B_0\ge 0$.

Finally, these constraints are projected onto the space of the galaxy-ISW angular correlation function for comparison with the data.  We do not attempt here to include the correlation
function data in the likelihood. That would require a joint re-analysis of the various data sets to capture the
covariance between the measurements.    It would also require an assessment of systematic errors in the correlation
due to uncertainties in the selection functions and other effects.  Instead we compare predictions
with the individual measurements and their errors as quoted in the literature.

\begin{figure}[htbp]
  \begin{center}
  \epsfxsize=3.3truein
    \epsffile{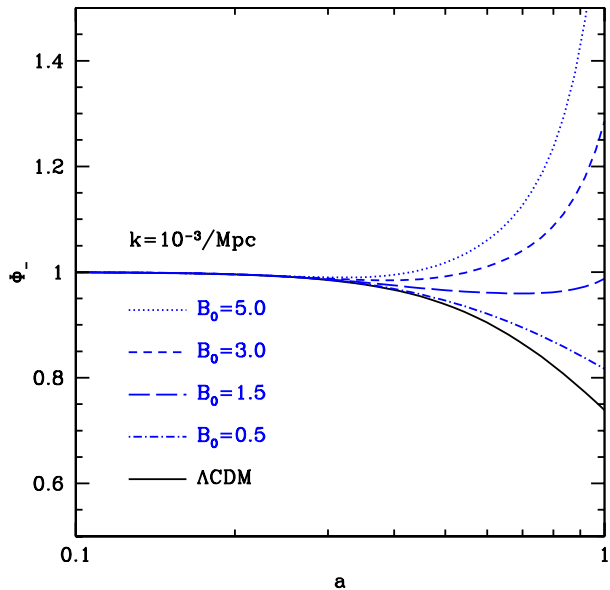}
    \caption{\footnotesize Time evolution of the
     effective gravitational potential $\Phi_- = (\Phi-\Psi)/2$ 
    for $k=10^{-3}/$Mpc near the peak contribution to the ISW effect at low multipoles.
    As the Compton wavelength parameter $B_0$ increases, the decay of the gravitational
    potential in $\Lambda$CDM decreases and eventually  turns to growth.   This reversal
    changes the sign of the ISW effect in overdense regions traced by galaxies.
}
\label{fig:pot}
\end{center}
\end{figure}

\section{Cosmological Constraints}
\label{sec:constraints}

\subsection{CMB}
\label{sec:CMB}

In $f(R)$ models that follow the $\Lambda$CDM expansion history considered here,
none of the CMB phenomenology at recombination is affected by the modification
to gravity.  The Compton wavelength parameter $B$ is
driven rapidly to
zero at high density and curvature.  Hence all of the successes of $\Lambda$CDM
in explaining the acoustic peaks carries over to these $f(R)$ models.

The impact of $f(R)$ gravity on the CMB comes exclusively through the so-called
Integrated Sachs-Wolfe (ISW) effect at the lowest multipole moments.  
The ISW effect arises from an imbalance between the blueshift
a CMB photon suffers while falling into a gravitational potential well and the
redshift  while climbing out if the gravitational potential evolves during transit.
With a cosmological constant, gravitational potentials decay during the acceleration
epoch.  For $f(R)$ models, the enhancement of the growth rate below the Compton
scale can change the decay into growth (see Fig.~\ref{fig:pot}).  This reversal
changes  the sign of the ISW effect.  CMB photons then become colder along directions
associated with overdense regions.

\begin{figure}[tb]
  \begin{center}
  \epsfxsize=3.3truein
    \epsffile{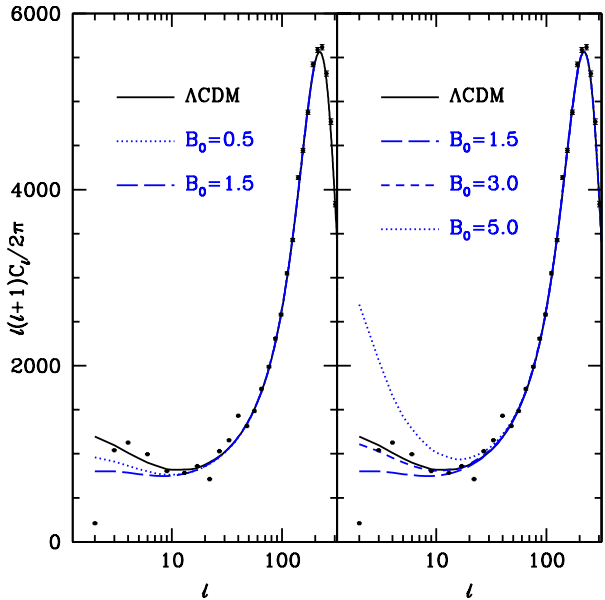}
    \caption{\footnotesize CMB angular power spectrum $C_{\ell}$ for $f(R)$ models
    with the Compton wavelength parameter
    $B_{0}=0$ ($\Lambda$CDM), 0.5, 1.5, 3.0, 5.0.  As $B_{0}$ increases, the
    ISW contributions to the low multipoles decrease, change sign, and then increase.
    WMAP3 data with noise error bars are overplotted and rule out $B_{0} \ge 4.3$ (95\% CL).
    }
\label{fig:cl}
\end{center}
\end{figure}

\begin{figure}[tb]
  \begin{center}
  \epsfxsize=3.3truein
    \epsffile{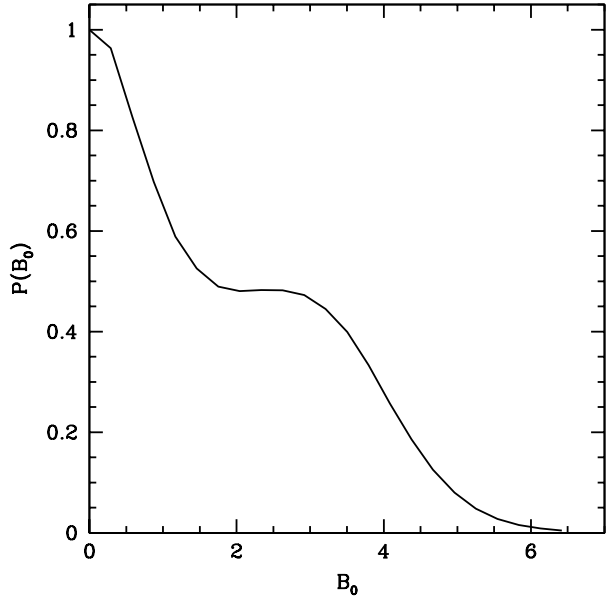}
    \caption{\footnotesize Posterior probability distribution for the Compton wavelength parameter
    $B_{0}$ inferred from the joint likelihood analysis of WMAP CMB angular power
    spectrum, SDSS LRG galaxy power spectrum, and SNLS supernovae data.  The upper 
    limit on $B_{0}$ from the joint constraint is driven by the CMB data, specifically the ISW effect.
}
\label{fig:like}
\end{center}
\end{figure}

Fig.~\ref{fig:cl}  illustrates the impact  of this effect on the CMB power spectrum.  
 As the Compton wavelength 
approaches the Gpc ($k \sim 10^{-3}$ Mpc$^{-1}$) scales associated with the low multipoles
of the ISW effect, the reduction in
the decay of the potential also suppresses the ISW effect.   Near $B_{0} \approx 1.5$ the
ISW effect is almost entirely absent at the quadrupole. 
Reduction in the amplitude of the quadrupole is in fact 
 weakly favored by the data but the large cosmic variance of the low multipoles
prevents this from being a significant improvement.  Moreover, the elimination of the
ISW effect at higher multipoles where the observed power is higher
 counteracts this improvement.
For $B_{0}\ga 1.5$  the Compton wavelength exceeds the scales
of interest and potential decay turns to potential growth.  By $B_{0}\sim 3$
the ISW effect has an equal amplitude to that of $\Lambda$CDM.  For
$B_{0}\ga 5$, it is too large to accommodate the WMAP data.   These
features drive the overall joint constraint on $B_{0}$ shown in Fig.~\ref{fig:like}.
The WMAP data also serve to fix the parameters that control the high redshift
universe.  In particular, it constrains the initial amplitude of power on scales
that are observed in galaxy surveys such as the SDSS LRG survey.

\subsection{Galaxy Power Spectrum}
\label{sec:Pk}

The time-dependent Compton wavelength of 
$f(R)$ models induces a more dramatic effect in the matter power spectrum during
the acceleration epoch.   In fact, average Compton wavelengths down to a few Mpc
are potentially observable in the linear power spectrum.  Under this scale, the enhanced
growth rate leads to excess power relative to the same initial power spectrum determined
by the WMAP data.   Unfortunately the overall change in 
power is degenerate with the unknown galaxy bias.   However if the Compton wavelength
appears between the few to 100 Mpc scale the distortion of the shape of the power spectrum is
potentially distinguishable in current surveys.  
In our parameterization this occurs for $B_{0} \sim 10^{-5}-10^{-2}$.

\begin{figure}[tb]
  \begin{center}
  \epsfxsize=3.3truein
    \epsffile{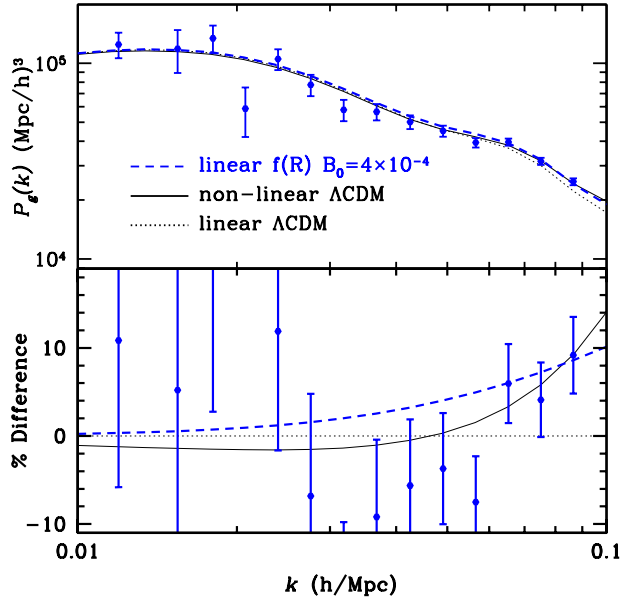}
    \caption{\footnotesize SDSS LRG galaxy power spectrum data compared with a linear
    $\Lambda$CDM power spectrum (see text), a linear $f(R)$ model with $B_{0}=4\times 10^{{-4}}$
    and the best fit nonlinear $\Lambda$CDM power spectrum.  While the data are better
    fit by $f(R)$ with {\it linear} power spectra, nonlinearities make $\Lambda$CDM a better
    fit to the data and open up a degeneracy between $f(R)$ modifications parameterized
    by $B_{0}$ and the non-linearity parameter $Q_{\rm nl}$.
}
\label{fig:Pk}
\end{center}
\end{figure}

The SDSS LRG data set in fact favor enhanced power over the {\it linear} $\Lambda$CDM
power spectrum.  In Fig.~\ref{fig:Pk}, we compare a linear $\Lambda$CDM
power spectrum
$(\omega_b=0.025, \omega_m=0.128, H_0=73, 10^{10}A_{s}=21.2, n_s=0.95, \tau=0.09, B_0=0, b=2.2$
and a linear $f(R)$ power spectrum with the same parameters and
$B_{0}=4\times 10^{-4}$.  The $f(R)$ model is in fact a  better fit to the shape of the
power spectrum, with $\Delta\chi^{2}_{\rm eff}=2\Delta \ln L \approx 11.5$ for this specific choice of bias.

\begin{figure}[tb]
  \begin{center}
  \epsfxsize=3.3truein
    \epsffile{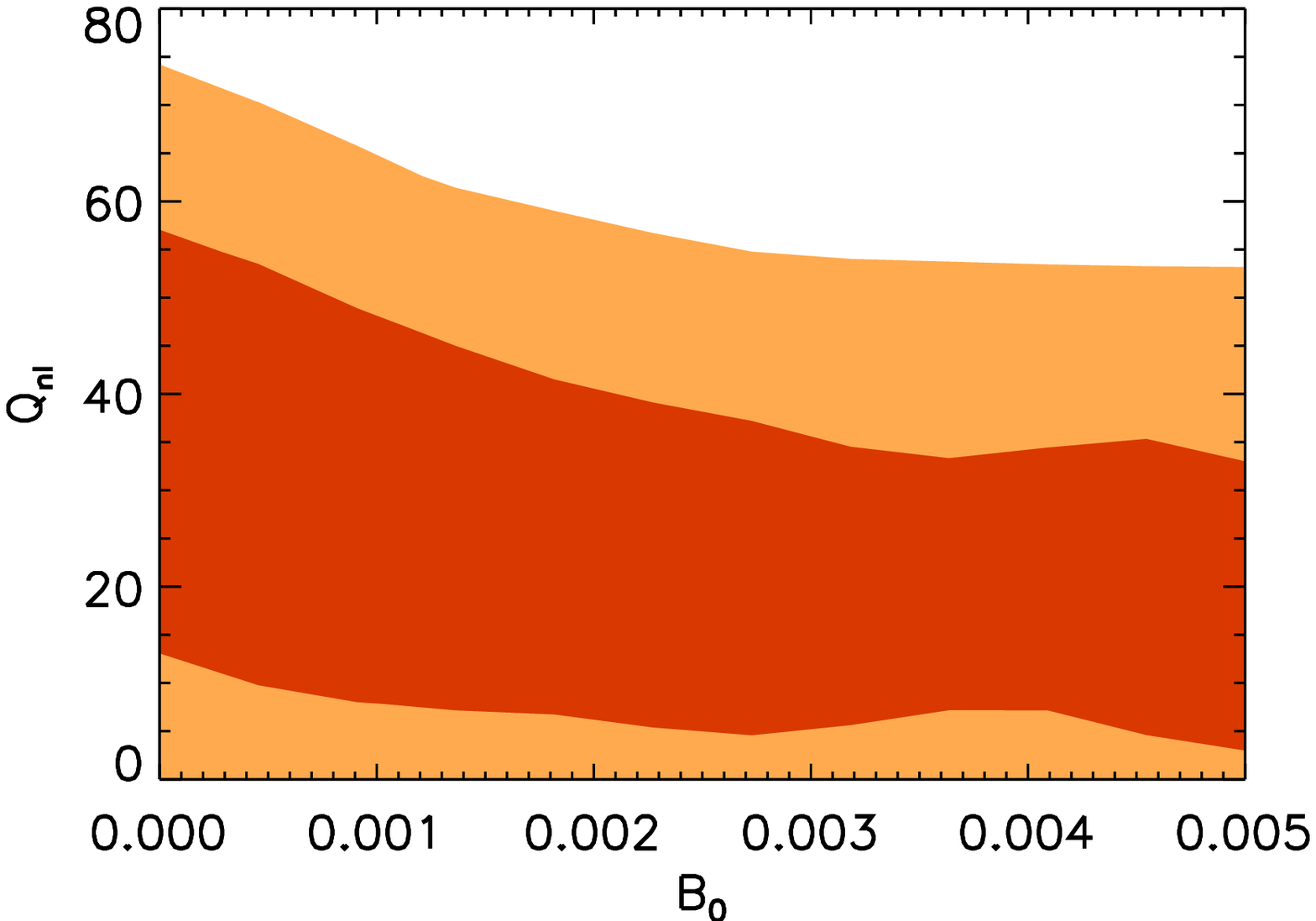}
     \epsfxsize=3.3truein
     \epsffile{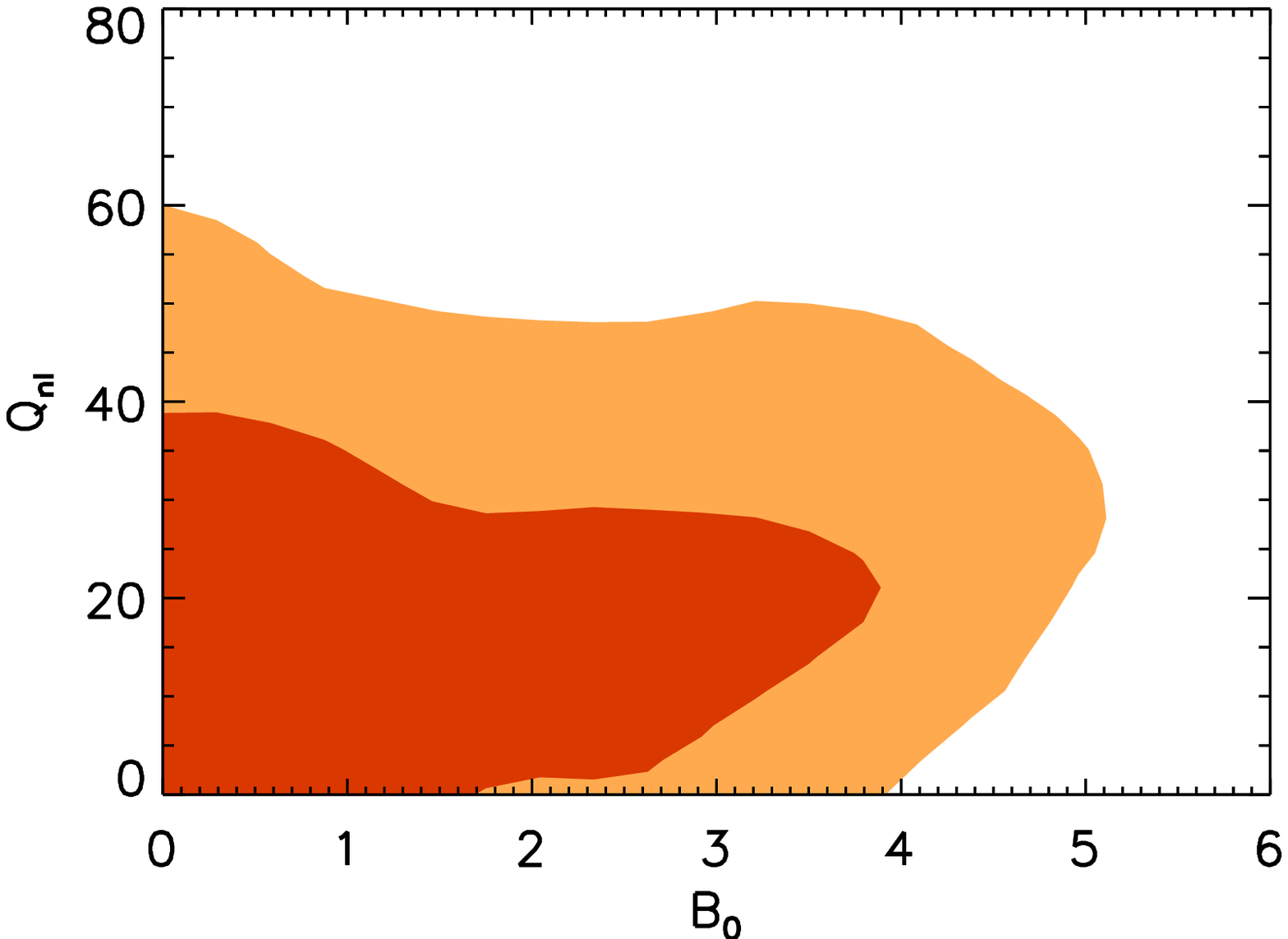}
    \caption{\footnotesize The 2D joint 68\% CL (dark) and 95\% CL (light) constraints on the Compton wavelength parameter $B_0$ and non-linearity parameter $Q_{\rm nl}$ inferred from the joint likelihood analysis of WMAP CMB angular power spectrum, SDSS LRG galaxy power spectrum, and SNLS supernovae data.  The upper panel focuses in on the range $B_0 < 0.005$ in order to highlight the 
anti-correlation between these two parameters discussed in the text, while the lower panel shows the full joint constraint.}
\label{fig:QnlB0}
\end{center}
\end{figure}

However in the $\Lambda$CDM model, we expect LRG galaxy clustering to be nonlinear
in exactly the region where these changes of shape occur.  In fact a non-linearity parameter 
of $Q_{\rm nl}=30$,
which represents a reasonable amount of non-linearity, produces an excellent fit to the
data (see Fig.~\ref{fig:Pk}). Hence,  $f(R)$ enhancements of small scale power are degenerate with non-linear
effects in $\Lambda$CDM and open up a corresponding degeneracy between 
$Q_{\rm nl}$ and $B_{0}$ (see  Fig.~\ref{fig:QnlB0}).    
The non-linear modification also introduces a small suppression
of power at intermediate $k$ for any value of $Q_{\rm nl}$ which also marginally improves
the fit.

\begin{figure}[tb]
  \begin{center}
  \epsfxsize=3.3truein
    \epsffile{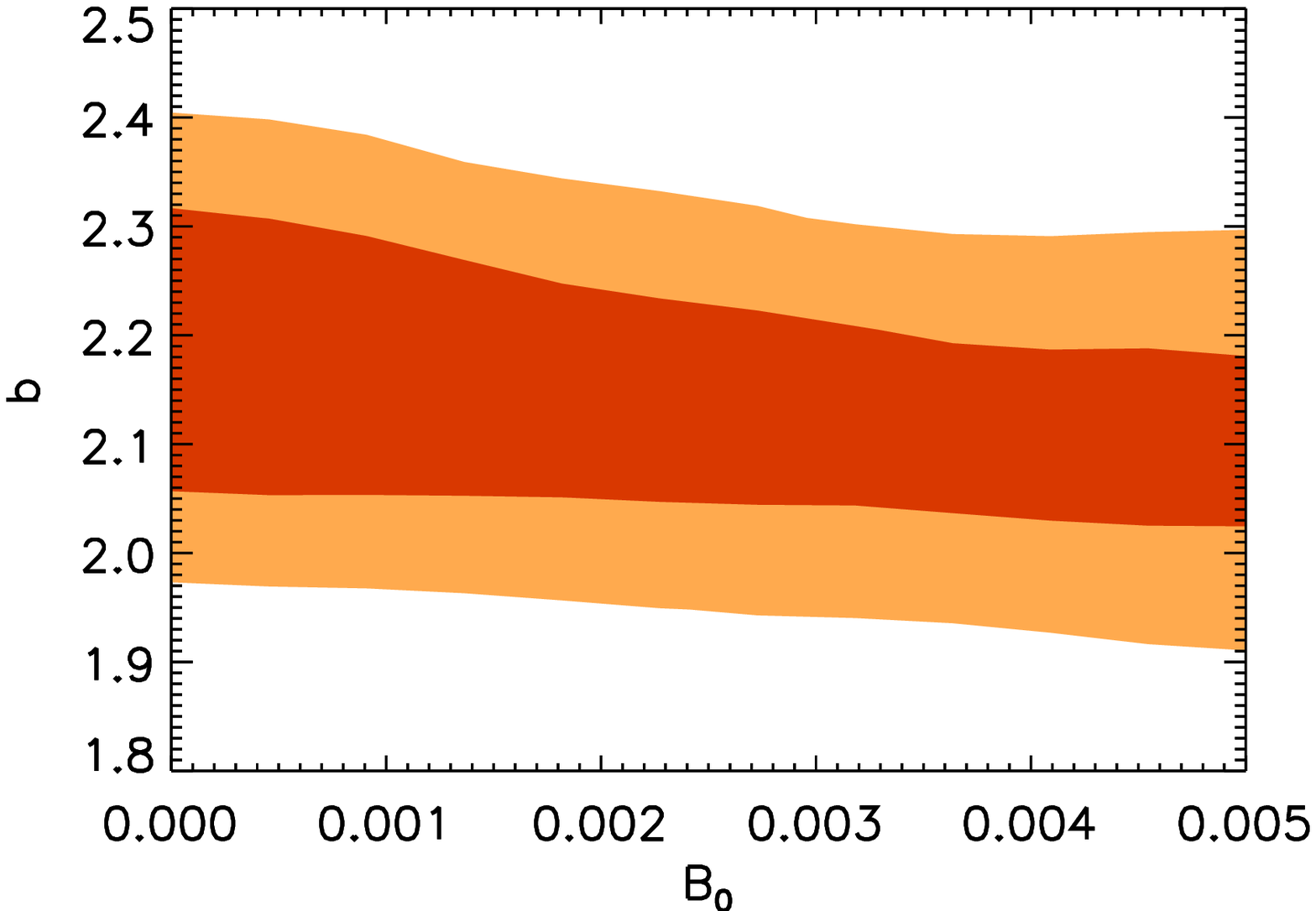}
     \epsfxsize=3.3truein
     \epsffile{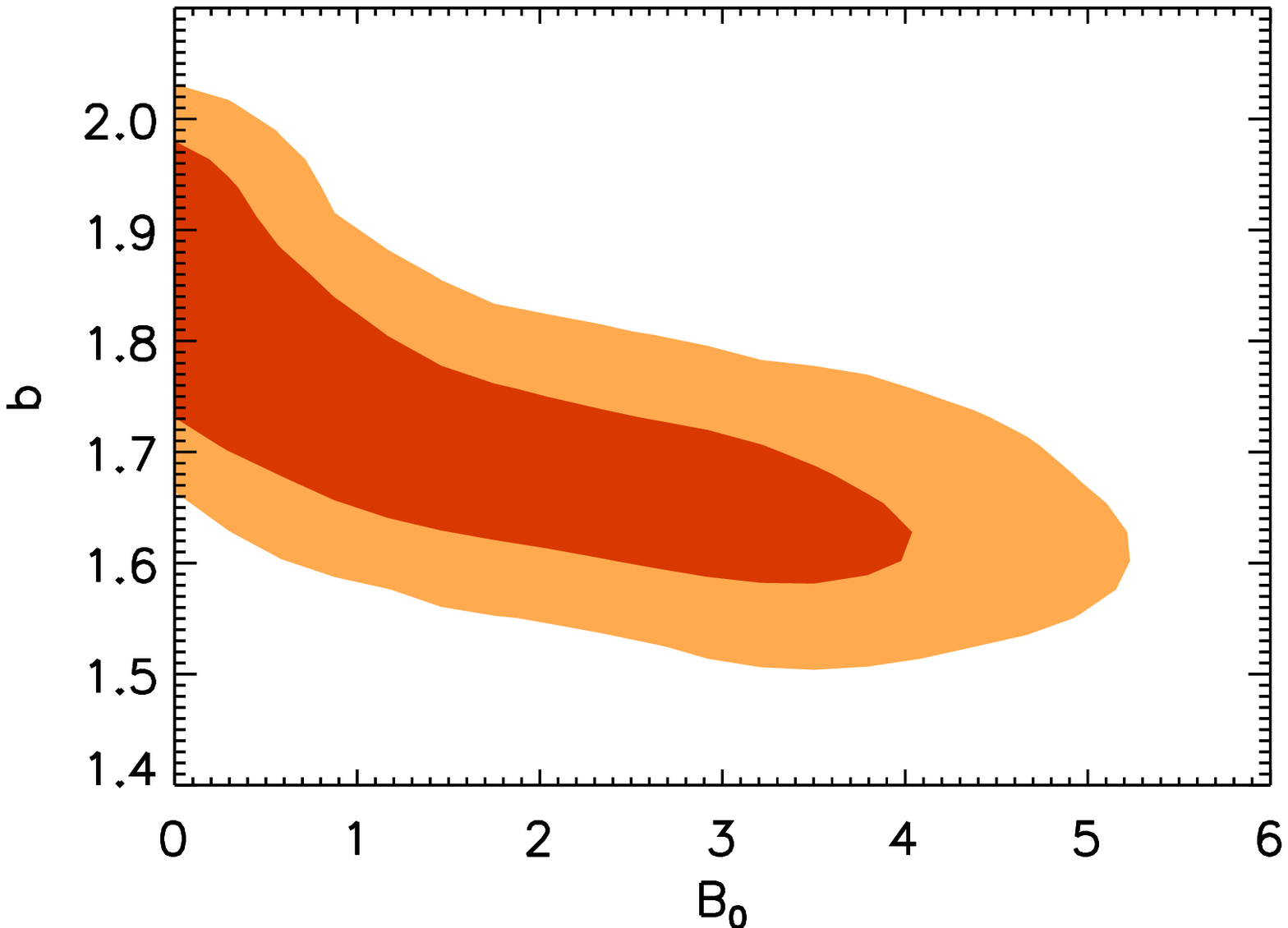}
    \caption{\footnotesize The 2D joint 68\% CL (dark) and 95\% CL (light) constraints on the Compton wavelength parameter $B_0$ and the linear galaxy bias $b$ inferred from the joint likelihood analysis of WMAP CMB angular power spectrum, SDSS LRG galaxy power spectrum, and SNLS supernovae data.  The upper panel focuses in on the range $B_0 < 0.005$ in order to highlight the 
anti-correlation between these two parameters discussed in the text, while the lower panel shows the full joint constraint.
}
\label{fig:bB0}
\end{center}
\end{figure}

Furthermore if $B_{0}\ga 10^{-2}$
the Compton scale
exceeds the largest scales in the survey at $k \sim 10^{-2} h/$Mpc.  The enhancement
of small scale power, while pronounced, becomes degenerate with the galaxy bias $b$.
With the primordial amplitude fixed by WMAP, the bias must decrease as $B_{0}$ increases,
leading to an anti-correlation between the two parameters (see Fig.~\ref{fig:bB0}).

Even order unity $B_0$ is allowed after marginalization.  The LRG data do weakly disfavor
even larger $B_0$ but the overall constraint is dominated by the CMB data
 (see Fig.~\ref{fig:like} and
compare $B_0=3$ relative to $B_0=0$,
 where the CMB 
spectra are nearly identical).   

It is the marginalization over
 the non-linear parameter $Q_{\rm nl}$ and bias which substantially degrades the ability
of the LRG data to constrain $f(R)$ models.  
This is a theoretical and not
an observational
limitation.  As we have seen, even the small value of $B_0 \sim 10^{-4}$ substantially
impacts the current data. 
 With cosmological simulations that address the non-linear evolution of the
matter power spectrum in $f(R)$ and the association of LRG's with dark matter haloes
in the simulation, this uncertainty can be lifted leading to substantially tighter bounds on
the Compton wavelength.

\begin{figure}[t]
  \begin{center}
  \epsfxsize=3.5truein
    \epsffile{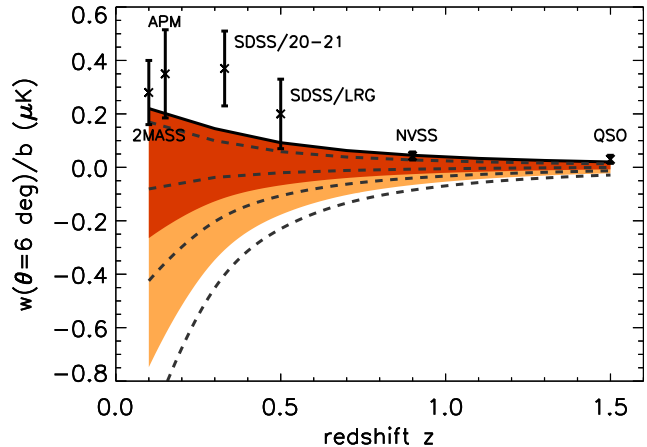} \hfill
    \caption{\footnotesize The galaxy-ISW angular correlation function at 6 degrees is shown as a function of redshift as measured by a variety of surveys (including quasars), with the bias divided out. The constraints on $f(R)$ models given by CMB, SDSS LRG galaxies and supernovae are projected onto this observable, with 68\% CL (dark) and 95\% CL (light) shaded regions shown. The $\Lambda$CDM limit of $B_0 \to 0$ is shown by a solid line, and the dashed lines (from top to bottom) correspond to $B_0=\{0.5, 1.5, 3.0, 5.0\}$.
}
\label{fig:wth}
\end{center}
\end{figure}

\subsection{Galaxy-ISW Correlation}
\label{sec:galISW}

The angular correlation between the CMB temperature field and galaxy number density field
induced by the ISW effect places an interesting constraint
on $f(R)$ models with averaged Compton wavelengths in the $>100$ Mpc regime.
As discussed in \ref{sec:CMB}, the ISW effect reverses sign for wavelengths
smaller than the average Compton 
wavelength during acceleration.    This reversal of sign causes galaxies to be 
anti-correlated with the CMB \cite{SonHuSaw06}.
For sufficiently large $B_{0}>1$,
the Compton wavelength
subtends a scale larger than the $1-10^\circ$ over which the correlation has
been measured.  In this regime, $f(R)$ models predict that galaxies are anti-correlated
with the CMB, in conflict with the data measured by 2MASS, APM, SDSS, NVSS and QSO
\cite{AfsLohStr04, BouCri04, FosGaz04, Giannantonio06, Cabre06}.

Figure~\ref{fig:wth} shows the angular correlation data at $6^\circ$ from a compilation
in \cite{Gaztanaga:2004sk} (updated by Gaztanaga, private communication).  
We compare these data with
predictions from models in the chain spanning the 68\% and 95\% allowed regions
given by the WMAP CMB, SDSS LRG power spectrum and supernovae data sets.   The
lack of anti-correlation in any given data point 
places an upper bound of $B_{0}\la  1$ at the
significance level of the detection.    This is approximately a factor of 4 improvement in $B_0$
over the other constraints or a factor of 2 in the Compton wavelength.
These individual constraints can be improved somewhat
by a full joint analysis of the correlation measurements but that lies beyond the scope of this
work.

\section{Discussion}
\label{sec:discussion}

We have analyzed current cosmological constraints on $f(R)$ acceleration
models from the
CMB, SDSS galaxy power spectrum, galaxy-ISW correlations and supernovae.
By choosing $f(R)$ models with a $\Lambda$CDM expansion history, we have
explored whether the unique signatures of the $f(R)$ modification to gravity
are seen in current data.

Despite the relatively large impact of $f(R)$ models on the matter power spectrum,
the strongest current constraints involve the modification these models induce
on the ISW effect in the CMB.    The growth of density perturbations is enhanced
under the Compton wavelength of the field $f_{R}=df/dR$  by scalar-tensor modifications
to the gravitational force law.  This enhancement turns the decay of
gravitational potentials during the acceleration epoch in $\Lambda$CDM into 
growth.   The joint constraint on the Compton wavelength parameter $B_{0} < 4.3$ (95\% CL),
from the WMAP CMB power spectrum, SDSS galaxy power spectrum and supernovae,
is driven by the CMB.  This constraint allows Compton wavelengths as large as
the current horizon and is a weak bound on $f(R)$ models.

A stronger bound of $B_{0} \la 1$ can be inferred from the positive correlation
between the CMB and a range of galaxy surveys from $z\sim 0.1-1.5$.  In $\Lambda$CDM
this positive correlation is induced by the ISW effect from a decaying potential.
In $f(R)$ models growing potentials convert the positive correlation into
anti-correlation in violation of the observations.   

The weak impact of the SDSS LRG galaxy power spectrum on our joint constraints
is due to a theoretical rather than observational limitation.  Intriguingly
the {\it linear} $f(R)$ power spectrum is a marginally better fit to the data than a
{\it linear} $\Lambda$CDM one.  However the nonlinear corrections expected
in $\Lambda$CDM bring the model back in agreement with the data. 
$N$-body simulations
have yet to determine the non-linear matter power spectrum and dark matter
halo (and hence galaxy) correlations in $f(R)$ models.  The likely impact of
non-linear evolution is degenerate with the enhancement of small scale linear
power seen in $f(R)$ models.  When non-linearity of the form expected for
$\Lambda$CDM is marginalized for $f(R)$ models, 
the impact of the galaxy power spectrum on joint constraints becomes very weak,
and Compton wavelengths out to the horizon length are allowed.   

In the future, the window between the horizon length and a few tens of Mpc
can be tested by larger photometric surveys that probe the shape of the
power spectrum across the whole range of
scales.  In principle, Compton wavelengths down to a few Mpc can be tested by
galaxy power spectra and cosmic shear from comparison of data with 
 simulations of $f(R)$ models.  

\smallskip
{\it Acknowledgments:} We thank Ignacy Sawicki for help during the initial phases of this work, 
Enrique Gaztanaga for the compilation of galaxy-ISW correlation data, and Marilena LoVerde for useful discussions.  
YS and WH are supported by the
U.S.~Dept. of Energy contract DE-FG02-90ER-40560. 
HVP is supported by NASA through Hubble Fellowship grant \#HF-01177.01-A awarded by the Space Telescope Science Institute, which is operated by the Association of Universities for Research in Astronomy, Inc., for NASA, under contract NAS 5-26555.  
WH is
additionally supported by the David and Lucile Packard Foundation.
This work was carried out at the KICP under NSF PHY-0114422.


\begin{thebibliography}{60}
\expandafter\ifx\csname natexlab\endcsname\relax\def\natexlab#1{#1}\fi
\expandafter\ifx\csname bibnamefont\endcsname\relax
  \def\bibnamefont#1{#1}\fi
\expandafter\ifx\csname bibfnamefont\endcsname\relax
  \def\bibfnamefont#1{#1}\fi
\expandafter\ifx\csname citenamefont\endcsname\relax
  \def\citenamefont#1{#1}\fi
\expandafter\ifx\csname url\endcsname\relax
  \def\url#1{\texttt{#1}}\fi
\expandafter\ifx\csname urlprefix\endcsname\relax\def\urlprefix{URL }\fi
\providecommand{\bibinfo}[2]{#2}
\providecommand{\eprint}[2][]{\url{#2}}

\bibitem[{\citenamefont{Starobinsky}(1980)}]{Sta80}
\bibinfo{author}{\bibfnamefont{A.~A.} \bibnamefont{Starobinsky}},
  \bibinfo{journal}{Phys. Lett.} \textbf{\bibinfo{volume}{B91}},
  \bibinfo{pages}{99} (\bibinfo{year}{1980}).

\bibitem[{\citenamefont{Carroll et~al.}(2004)\citenamefont{Carroll, Duvvuri,
  Trodden, and Turner}}]{Caretal03}
\bibinfo{author}{\bibfnamefont{S.~M.} \bibnamefont{Carroll}},
  \bibinfo{author}{\bibfnamefont{V.}~\bibnamefont{Duvvuri}},
  \bibinfo{author}{\bibfnamefont{M.}~\bibnamefont{Trodden}}, \bibnamefont{and}
  \bibinfo{author}{\bibfnamefont{M.~S.} \bibnamefont{Turner}},
  \bibinfo{journal}{Phys. Rev.} \textbf{\bibinfo{volume}{D70}},
  \bibinfo{pages}{043528} (\bibinfo{year}{2004}), \eprint{astro-ph/0306438}.

\bibitem[{\citenamefont{Capozziello et~al.}(2003)\citenamefont{Capozziello,
  Carloni, and Troisi}}]{CapCarTro03}
\bibinfo{author}{\bibfnamefont{S.}~\bibnamefont{Capozziello}},
  \bibinfo{author}{\bibfnamefont{S.}~\bibnamefont{Carloni}}, \bibnamefont{and}
  \bibinfo{author}{\bibfnamefont{A.}~\bibnamefont{Troisi}}
  (\bibinfo{year}{2003}), \eprint{astro-ph/0303041}.

\bibitem[{\citenamefont{Nojiri and Odintsov}(2003)}]{NojOdi03}
\bibinfo{author}{\bibfnamefont{S.}~\bibnamefont{Nojiri}} \bibnamefont{and}
  \bibinfo{author}{\bibfnamefont{S.~D.} \bibnamefont{Odintsov}},
  \bibinfo{journal}{Phys. Rev.} \textbf{\bibinfo{volume}{D68}},
  \bibinfo{pages}{123512} (\bibinfo{year}{2003}), \eprint{hep-th/0307288}.

\bibitem[{\citenamefont{Poplawski}(2006)}]{Poplawski:2006kv}
\bibinfo{author}{\bibfnamefont{N.~J.} \bibnamefont{Poplawski}},
  \bibinfo{journal}{Phys. Rev.} \textbf{\bibinfo{volume}{D74}},
  \bibinfo{pages}{084032} (\bibinfo{year}{2006}), \eprint{gr-qc/0607124}.

\bibitem[{\citenamefont{de~la Cruz-Dombriz and
  Dobado}(2006)}]{delaCruz-Dombriz:2006fj}
\bibinfo{author}{\bibfnamefont{A.}~\bibnamefont{de~la Cruz-Dombriz}}
  \bibnamefont{and} \bibinfo{author}{\bibfnamefont{A.}~\bibnamefont{Dobado}},
  \bibinfo{journal}{Phys. Rev.} \textbf{\bibinfo{volume}{D74}},
  \bibinfo{pages}{087501} (\bibinfo{year}{2006}), \eprint{gr-qc/0607118}.

\bibitem[{\citenamefont{Brookfield et~al.}(2006)\citenamefont{Brookfield,
  van~de Bruck, and Hall}}]{Brookfield:2006mq}
\bibinfo{author}{\bibfnamefont{A.~W.} \bibnamefont{Brookfield}},
  \bibinfo{author}{\bibfnamefont{C.}~\bibnamefont{van~de Bruck}},
  \bibnamefont{and} \bibinfo{author}{\bibfnamefont{L.~M.~H.}
  \bibnamefont{Hall}}, \bibinfo{journal}{Phys. Rev.}
  \textbf{\bibinfo{volume}{D74}}, \bibinfo{pages}{064028}
  (\bibinfo{year}{2006}), \eprint{hep-th/0608015}.

\bibitem[{\citenamefont{Sotiriou}(2007)}]{Sotiriou:2006sf}
\bibinfo{author}{\bibfnamefont{T.~P.} \bibnamefont{Sotiriou}},
  \bibinfo{journal}{Phys. Lett.} \textbf{\bibinfo{volume}{B645}},
  \bibinfo{pages}{389} (\bibinfo{year}{2007}), \eprint{gr-qc/0611107}.

\bibitem[{\citenamefont{Sotiriou}(2006)}]{Sotiriou:2006hs}
\bibinfo{author}{\bibfnamefont{T.~P.} \bibnamefont{Sotiriou}},
  \bibinfo{journal}{Class. Quant. Grav.} \textbf{\bibinfo{volume}{23}},
  \bibinfo{pages}{5117} (\bibinfo{year}{2006}), \eprint{gr-qc/0604028}.

\bibitem[{\citenamefont{Bean et~al.}(2007)\citenamefont{Bean, Bernat, Pogosian,
  Silvestri, and Trodden}}]{Bean:2006up}
\bibinfo{author}{\bibfnamefont{R.}~\bibnamefont{Bean}},
  \bibinfo{author}{\bibfnamefont{D.}~\bibnamefont{Bernat}},
  \bibinfo{author}{\bibfnamefont{L.}~\bibnamefont{Pogosian}},
  \bibinfo{author}{\bibfnamefont{A.}~\bibnamefont{Silvestri}},
  \bibnamefont{and} \bibinfo{author}{\bibfnamefont{M.}~\bibnamefont{Trodden}},
  \bibinfo{journal}{Phys. Rev.} \textbf{\bibinfo{volume}{D75}},
  \bibinfo{pages}{064020} (\bibinfo{year}{2007}), \eprint{astro-ph/0611321}.

\bibitem[{\citenamefont{Amendola et~al.}(2007)\citenamefont{Amendola, Gannouji,
  Polarski, and Tsujikawa}}]{Amendola:2006we}
\bibinfo{author}{\bibfnamefont{L.}~\bibnamefont{Amendola}},
  \bibinfo{author}{\bibfnamefont{R.}~\bibnamefont{Gannouji}},
  \bibinfo{author}{\bibfnamefont{D.}~\bibnamefont{Polarski}}, \bibnamefont{and}
  \bibinfo{author}{\bibfnamefont{S.}~\bibnamefont{Tsujikawa}},
  \bibinfo{journal}{Phys. Rev.} \textbf{\bibinfo{volume}{D75}},
  \bibinfo{pages}{083504} (\bibinfo{year}{2007}), \eprint{gr-qc/0612180}.

\bibitem[{\citenamefont{Baghram et~al.}(2007)\citenamefont{Baghram, Farhang,
  and Rahvar}}]{Baghram:2007df}
\bibinfo{author}{\bibfnamefont{S.}~\bibnamefont{Baghram}},
  \bibinfo{author}{\bibfnamefont{M.}~\bibnamefont{Farhang}}, \bibnamefont{and}
  \bibinfo{author}{\bibfnamefont{S.}~\bibnamefont{Rahvar}},
  \bibinfo{journal}{Phys. Rev.} \textbf{\bibinfo{volume}{D75}},
  \bibinfo{pages}{044024} (\bibinfo{year}{2007}), \eprint{astro-ph/0701013}.

\bibitem[{\citenamefont{Bazeia et~al.}(2007)\citenamefont{Bazeia, Carneiro~da
  Cunha, Menezes, and Petrov}}]{Bazeia:2007jj}
\bibinfo{author}{\bibfnamefont{D.}~\bibnamefont{Bazeia}},
  \bibinfo{author}{\bibfnamefont{B.}~\bibnamefont{Carneiro~da Cunha}},
  \bibinfo{author}{\bibfnamefont{R.}~\bibnamefont{Menezes}}, \bibnamefont{and}
  \bibinfo{author}{\bibfnamefont{A.~Y.} \bibnamefont{Petrov}}
  (\bibinfo{year}{2007}), \eprint{hep-th/0701106}.

\bibitem[{\citenamefont{Li and Barrow}(2007)}]{Li:2007xn}
\bibinfo{author}{\bibfnamefont{B.}~\bibnamefont{Li}} \bibnamefont{and}
  \bibinfo{author}{\bibfnamefont{J.~D.} \bibnamefont{Barrow}},
  \bibinfo{journal}{Phys. Rev.} \textbf{\bibinfo{volume}{D75}},
  \bibinfo{pages}{084010} (\bibinfo{year}{2007}), \eprint{gr-qc/0701111}.

\bibitem[{\citenamefont{Bludman}(2007)}]{Bludman:2007kg}
\bibinfo{author}{\bibfnamefont{S.}~\bibnamefont{Bludman}}
  (\bibinfo{year}{2007}), \eprint{astro-ph/0702085}.

\bibitem[{\citenamefont{Rador}(2007)}]{Rador:2007wq}
\bibinfo{author}{\bibfnamefont{T.}~\bibnamefont{Rador}} (\bibinfo{year}{2007}),
  \eprint{hep-th/0702081}.

\bibitem[{\citenamefont{Faraoni}(2006)}]{Faraoni:2006sy}
\bibinfo{author}{\bibfnamefont{V.}~\bibnamefont{Faraoni}},
  \bibinfo{journal}{Phys. Rev.} \textbf{\bibinfo{volume}{D74}},
  \bibinfo{pages}{104017} (\bibinfo{year}{2006}), \eprint{astro-ph/0610734}.

\bibitem[{\citenamefont{Koivisto}(2006)}]{Koivisto:2006ie}
\bibinfo{author}{\bibfnamefont{T.}~\bibnamefont{Koivisto}},
  \bibinfo{journal}{Phys. Rev.} \textbf{\bibinfo{volume}{D73}},
  \bibinfo{pages}{083517} (\bibinfo{year}{2006}), \eprint{astro-ph/0602031}.

\bibitem[{\citenamefont{Capozziello and Garattini}(2007)}]{Capozziello:2007gm}
\bibinfo{author}{\bibfnamefont{S.}~\bibnamefont{Capozziello}} \bibnamefont{and}
  \bibinfo{author}{\bibfnamefont{R.}~\bibnamefont{Garattini}},
  \bibinfo{journal}{Class. Quant. Grav.} \textbf{\bibinfo{volume}{24}},
  \bibinfo{pages}{1627} (\bibinfo{year}{2007}), \eprint{gr-qc/0702075}.

\bibitem[{\citenamefont{Nojiri and
  Odintsov}(2006{\natexlab{a}})}]{Nojiri:2006gh}
\bibinfo{author}{\bibfnamefont{S.}~\bibnamefont{Nojiri}} \bibnamefont{and}
  \bibinfo{author}{\bibfnamefont{S.~D.} \bibnamefont{Odintsov}},
  \bibinfo{journal}{Phys. Rev.} \textbf{\bibinfo{volume}{D74}},
  \bibinfo{pages}{086005} (\bibinfo{year}{2006}{\natexlab{a}}),
  \eprint{hep-th/0608008}.

\bibitem[{\citenamefont{Nojiri and
  Odintsov}(2006{\natexlab{b}})}]{Nojiri:2006su}
\bibinfo{author}{\bibfnamefont{S.}~\bibnamefont{Nojiri}} \bibnamefont{and}
  \bibinfo{author}{\bibfnamefont{S.~D.} \bibnamefont{Odintsov}}
  (\bibinfo{year}{2006}{\natexlab{b}}), \eprint{hep-th/0610164}.

\bibitem[{\citenamefont{Capozziello
  et~al.}(2006{\natexlab{a}})\citenamefont{Capozziello, Nojiri, Odintsov, and
  Troisi}}]{Capozziello:2006dj}
\bibinfo{author}{\bibfnamefont{S.}~\bibnamefont{Capozziello}},
  \bibinfo{author}{\bibfnamefont{S.}~\bibnamefont{Nojiri}},
  \bibinfo{author}{\bibfnamefont{S.~D.} \bibnamefont{Odintsov}},
  \bibnamefont{and} \bibinfo{author}{\bibfnamefont{A.}~\bibnamefont{Troisi}},
  \bibinfo{journal}{Phys. Lett.} \textbf{\bibinfo{volume}{B639}},
  \bibinfo{pages}{135} (\bibinfo{year}{2006}{\natexlab{a}}),
  \eprint{astro-ph/0604431}.

\bibitem[{\citenamefont{Fay et~al.}(2007)\citenamefont{Fay, Nesseris, and
  Perivolaropoulos}}]{Fay:2007uy}
\bibinfo{author}{\bibfnamefont{S.}~\bibnamefont{Fay}},
  \bibinfo{author}{\bibfnamefont{S.}~\bibnamefont{Nesseris}}, \bibnamefont{and}
  \bibinfo{author}{\bibfnamefont{L.}~\bibnamefont{Perivolaropoulos}}
  (\bibinfo{year}{2007}), \eprint{gr-qc/0703006}.

\bibitem[{\citenamefont{Amendola and Tsujikawa}(2007)}]{Amendola:2007nt}
\bibinfo{author}{\bibfnamefont{L.}~\bibnamefont{Amendola}} \bibnamefont{and}
  \bibinfo{author}{\bibfnamefont{S.}~\bibnamefont{Tsujikawa}}
  (\bibinfo{year}{2007}), \eprint{arXiv:0705.0396 [astro-ph]}.

\bibitem[{\citenamefont{Song et~al.}(2007)\citenamefont{Song, Hu, and
  Sawicki}}]{SonHuSaw06}
\bibinfo{author}{\bibfnamefont{Y.-S.} \bibnamefont{Song}},
  \bibinfo{author}{\bibfnamefont{W.}~\bibnamefont{Hu}}, \bibnamefont{and}
  \bibinfo{author}{\bibfnamefont{I.}~\bibnamefont{Sawicki}},
  \bibinfo{journal}{Phys. Rev.} \textbf{\bibinfo{volume}{D75}},
  \bibinfo{pages}{044004} (\bibinfo{year}{2007}), \eprint{astro-ph/0610532}.

\bibitem[{\citenamefont{{Sawicki} and {Hu}}(2007)}]{SawHu07}
\bibinfo{author}{\bibfnamefont{I.}~\bibnamefont{{Sawicki}}} \bibnamefont{and}
  \bibinfo{author}{\bibfnamefont{W.}~\bibnamefont{{Hu}}},
  \bibinfo{journal}{\prd} \textbf{\bibinfo{volume}{{\rm in press}}},
  \bibinfo{pages}{astro} (\bibinfo{year}{2007}), \eprint{astro-ph/0702278}.

\bibitem[{\citenamefont{Tsujikawa}(2007)}]{Tsujikawa:2007gd}
\bibinfo{author}{\bibfnamefont{S.}~\bibnamefont{Tsujikawa}}
  (\bibinfo{year}{2007}), \eprint{arXiv:0705.1032 [astro-ph]}.

\bibitem[{\citenamefont{Appleby and Battye}(2007)}]{Appleby:2007vb}
\bibinfo{author}{\bibfnamefont{S.~A.} \bibnamefont{Appleby}} \bibnamefont{and}
  \bibinfo{author}{\bibfnamefont{R.~A.} \bibnamefont{Battye}}
  (\bibinfo{year}{2007}), \eprint{arXiv:0705.3199 [astro-ph]}.

\bibitem[{\citenamefont{Charmousis et~al.}(2007)\citenamefont{Charmousis,
  Gregory, and Padilla}}]{Charmousis:2007ji}
\bibinfo{author}{\bibfnamefont{C.}~\bibnamefont{Charmousis}},
  \bibinfo{author}{\bibfnamefont{R.}~\bibnamefont{Gregory}}, \bibnamefont{and}
  \bibinfo{author}{\bibfnamefont{A.}~\bibnamefont{Padilla}}
  (\bibinfo{year}{2007}), \eprint{arXiv:0706.0857 [hep-th]}.

\bibitem[{\citenamefont{De~Felice et~al.}(2007)\citenamefont{De~Felice,
  Mukherjee, and Wang}}]{DeFelice:2007ez}
\bibinfo{author}{\bibfnamefont{A.}~\bibnamefont{De~Felice}},
  \bibinfo{author}{\bibfnamefont{P.}~\bibnamefont{Mukherjee}},
  \bibnamefont{and} \bibinfo{author}{\bibfnamefont{Y.}~\bibnamefont{Wang}}
  (\bibinfo{year}{2007}), \eprint{arXiv:0706.1197 [astro-ph]}.

\bibitem[{\citenamefont{Chiba}(2003)}]{Chi03}
\bibinfo{author}{\bibfnamefont{T.}~\bibnamefont{Chiba}},
  \bibinfo{journal}{Phys. Lett.} \textbf{\bibinfo{volume}{B575}},
  \bibinfo{pages}{1} (\bibinfo{year}{2003}), \eprint{astro-ph/0307338}.

\bibitem[{\citenamefont{Chiba et~al.}(2006)\citenamefont{Chiba, Smith, and
  Erickcek}}]{Chiba:2006jp}
\bibinfo{author}{\bibfnamefont{T.}~\bibnamefont{Chiba}},
  \bibinfo{author}{\bibfnamefont{T.~L.} \bibnamefont{Smith}}, \bibnamefont{and}
  \bibinfo{author}{\bibfnamefont{A.~L.} \bibnamefont{Erickcek}}
  (\bibinfo{year}{2006}), \eprint{astro-ph/0611867}.

\bibitem[{\citenamefont{Erickcek et~al.}(2006)\citenamefont{Erickcek, Smith,
  and Kamionkowski}}]{EriSmiKam06}
\bibinfo{author}{\bibfnamefont{A.~L.} \bibnamefont{Erickcek}},
  \bibinfo{author}{\bibfnamefont{T.~L.} \bibnamefont{Smith}}, \bibnamefont{and}
  \bibinfo{author}{\bibfnamefont{M.}~\bibnamefont{Kamionkowski}},
  \bibinfo{journal}{Phys. Rev.} \textbf{\bibinfo{volume}{D74}},
  \bibinfo{pages}{121501} (\bibinfo{year}{2006}), \eprint{astro-ph/0610483}.

\bibitem[{\citenamefont{Khoury and Weltman}(2004)}]{KhoWel04}
\bibinfo{author}{\bibfnamefont{J.}~\bibnamefont{Khoury}} \bibnamefont{and}
  \bibinfo{author}{\bibfnamefont{A.}~\bibnamefont{Weltman}},
  \bibinfo{journal}{Phys. Rev.} \textbf{\bibinfo{volume}{D69}},
  \bibinfo{pages}{044026} (\bibinfo{year}{2004}), \eprint{astro-ph/0309411}.

\bibitem[{\citenamefont{Mota and Barrow}(2004)}]{MotBar04}
\bibinfo{author}{\bibfnamefont{D.~F.} \bibnamefont{Mota}} \bibnamefont{and}
  \bibinfo{author}{\bibfnamefont{J.~D.} \bibnamefont{Barrow}},
  \bibinfo{journal}{Mon. Not. Roy. Astron. Soc.}
  \textbf{\bibinfo{volume}{349}}, \bibinfo{pages}{291} (\bibinfo{year}{2004}),
  \eprint{astro-ph/0309273}.

\bibitem[{\citenamefont{Navarro and Van~Acoleyen}(2006)}]{NavAco06}
\bibinfo{author}{\bibfnamefont{I.}~\bibnamefont{Navarro}} \bibnamefont{and}
  \bibinfo{author}{\bibfnamefont{K.}~\bibnamefont{Van~Acoleyen}}
  (\bibinfo{year}{2006}), \eprint{gr-qc/0611127}.

\bibitem[{\citenamefont{Faulkner et~al.}(2006)\citenamefont{Faulkner, Tegmark,
  Bunn, and Mao}}]{FauTegBun06}
\bibinfo{author}{\bibfnamefont{T.}~\bibnamefont{Faulkner}},
  \bibinfo{author}{\bibfnamefont{M.}~\bibnamefont{Tegmark}},
  \bibinfo{author}{\bibfnamefont{E.~F.} \bibnamefont{Bunn}}, \bibnamefont{and}
  \bibinfo{author}{\bibfnamefont{Y.}~\bibnamefont{Mao}} (\bibinfo{year}{2006}),
  \eprint{astro-ph/0612569}.

\bibitem[{\citenamefont{{Hu} and {Sawicki}}(2007)}]{HuSaw07}
\bibinfo{author}{\bibfnamefont{W.}~\bibnamefont{{Hu}}} \bibnamefont{and}
  \bibinfo{author}{\bibfnamefont{I.}~\bibnamefont{{Sawicki}}},
  \bibinfo{journal}{\prd} \textbf{\bibinfo{volume}{\rm submitted}}
  (\bibinfo{year}{2007}), \eprint{0705.1158 [astro-ph]}.

\bibitem[{\citenamefont{Spergel et~al.}(2006)}]{Speetal06}
\bibinfo{author}{\bibfnamefont{D.~N.} \bibnamefont{Spergel}}
  \bibnamefont{et~al.} (\bibinfo{year}{2006}), \eprint{astro-ph/0603449}.

\bibitem[{\citenamefont{Tegmark et~al.}(2006)}]{Tegetal06}
\bibinfo{author}{\bibfnamefont{M.}~\bibnamefont{Tegmark}} \bibnamefont{et~al.},
  \bibinfo{journal}{Phys. Rev.} \textbf{\bibinfo{volume}{D74}},
  \bibinfo{pages}{123507} (\bibinfo{year}{2006}), \eprint{astro-ph/0608632}.

\bibitem[{\citenamefont{Astier et~al.}(2006)}]{Astier}
\bibinfo{author}{\bibfnamefont{P.}~\bibnamefont{Astier}} \bibnamefont{et~al.},
  \bibinfo{journal}{Astron. Astrophys.} \textbf{\bibinfo{volume}{447}},
  \bibinfo{pages}{31} (\bibinfo{year}{2006}), \eprint{astro-ph/0510447}.

\bibitem[{\citenamefont{Afshordi et~al.}(2004)\citenamefont{Afshordi, Loh, and
  Strauss}}]{AfsLohStr04}
\bibinfo{author}{\bibfnamefont{N.}~\bibnamefont{Afshordi}},
  \bibinfo{author}{\bibfnamefont{Y.-S.} \bibnamefont{Loh}}, \bibnamefont{and}
  \bibinfo{author}{\bibfnamefont{M.~A.} \bibnamefont{Strauss}},
  \bibinfo{journal}{Phys. Rev.} \textbf{\bibinfo{volume}{D69}},
  \bibinfo{pages}{083524} (\bibinfo{year}{2004}), \eprint{astro-ph/0308260}.

\bibitem[{\citenamefont{{Boughn} and {Crittenden}}(2004)}]{BouCri04}
\bibinfo{author}{\bibfnamefont{S.}~\bibnamefont{{Boughn}}} \bibnamefont{and}
  \bibinfo{author}{\bibfnamefont{R.}~\bibnamefont{{Crittenden}}},
  \bibinfo{journal}{Nature} \textbf{\bibinfo{volume}{427}}, \bibinfo{pages}{45}
  (\bibinfo{year}{2004}).

\bibitem[{\citenamefont{{Fosalba} and {Gaztanaga}}(2004)}]{FosGaz04}
\bibinfo{author}{\bibfnamefont{P.}~\bibnamefont{{Fosalba}}} \bibnamefont{and}
  \bibinfo{author}{\bibfnamefont{E.}~\bibnamefont{{Gaztanaga}}},
  \bibinfo{journal}{\mnras} \textbf{\bibinfo{volume}{350}}, \bibinfo{pages}{37}
  (\bibinfo{year}{2004}).

\bibitem[{\citenamefont{Giannantonio et~al.}(2006)}]{Giannantonio06}
\bibinfo{author}{\bibfnamefont{T.}~\bibnamefont{Giannantonio}}
  \bibnamefont{et~al.}, \bibinfo{journal}{Phys. Rev.}
  \textbf{\bibinfo{volume}{D74}}, \bibinfo{pages}{063520}
  (\bibinfo{year}{2006}), \eprint{astro-ph/0607572}.

\bibitem[{\citenamefont{{Cabr{\'e}} et~al.}(2006)\citenamefont{{Cabr{\'e}},
  {Gazta{\~n}aga}, {Manera}, {Fosalba}, and {Castander}}}]{Cabre06}
\bibinfo{author}{\bibfnamefont{A.}~\bibnamefont{{Cabr{\'e}}}},
  \bibinfo{author}{\bibfnamefont{E.}~\bibnamefont{{Gazta{\~n}aga}}},
  \bibinfo{author}{\bibfnamefont{M.}~\bibnamefont{{Manera}}},
  \bibinfo{author}{\bibfnamefont{P.}~\bibnamefont{{Fosalba}}},
  \bibnamefont{and}
  \bibinfo{author}{\bibfnamefont{F.}~\bibnamefont{{Castander}}},
  \bibinfo{journal}{\mnras} \textbf{\bibinfo{volume}{372}},
  \bibinfo{pages}{L23} (\bibinfo{year}{2006}), \eprint{arXiv:astro-ph/0603690}.

\bibitem[{\citenamefont{Multamaki and Vilja}(2006)}]{Multamaki:2005zs}
\bibinfo{author}{\bibfnamefont{T.}~\bibnamefont{Multamaki}} \bibnamefont{and}
  \bibinfo{author}{\bibfnamefont{I.}~\bibnamefont{Vilja}},
  \bibinfo{journal}{Phys. Rev.} \textbf{\bibinfo{volume}{D73}},
  \bibinfo{pages}{024018} (\bibinfo{year}{2006}), \eprint{astro-ph/0506692}.

\bibitem[{\citenamefont{Capozziello
  et~al.}(2006{\natexlab{b}})\citenamefont{Capozziello, Nojiri, Odintsov, and
  Troisi}}]{CapNojOdiTro06}
\bibinfo{author}{\bibfnamefont{S.}~\bibnamefont{Capozziello}},
  \bibinfo{author}{\bibfnamefont{S.}~\bibnamefont{Nojiri}},
  \bibinfo{author}{\bibfnamefont{S.~D.} \bibnamefont{Odintsov}},
  \bibnamefont{and} \bibinfo{author}{\bibfnamefont{A.}~\bibnamefont{Troisi}},
  \bibinfo{journal}{Phys. Lett.} \textbf{\bibinfo{volume}{B639}},
  \bibinfo{pages}{135} (\bibinfo{year}{2006}{\natexlab{b}}),
  \eprint{astro-ph/0604431}.

\bibitem[{\citenamefont{Nojiri and Odintsov}(2006{\natexlab{c}})}]{NojOdi06}
\bibinfo{author}{\bibfnamefont{S.}~\bibnamefont{Nojiri}} \bibnamefont{and}
  \bibinfo{author}{\bibfnamefont{S.~D.} \bibnamefont{Odintsov}}
  (\bibinfo{year}{2006}{\natexlab{c}}), \eprint{hep-th/0611071}.

\bibitem[{\citenamefont{Zhang}(2006)}]{Zha05}
\bibinfo{author}{\bibfnamefont{P.}~\bibnamefont{Zhang}},
  \bibinfo{journal}{Phys. Rev.} \textbf{\bibinfo{volume}{D73}},
  \bibinfo{pages}{123504} (\bibinfo{year}{2006}), \eprint{astro-ph/0511218}.

\bibitem[{\citenamefont{{Lewis} et~al.}(2000)\citenamefont{{Lewis},
  {Challinor}, and {Lasenby}}}]{Lewetal00}
\bibinfo{author}{\bibfnamefont{A.}~\bibnamefont{{Lewis}}},
  \bibinfo{author}{\bibfnamefont{A.}~\bibnamefont{{Challinor}}},
  \bibnamefont{and}
  \bibinfo{author}{\bibfnamefont{A.}~\bibnamefont{{Lasenby}}},
  \bibinfo{journal}{\apj} \textbf{\bibinfo{volume}{538}}, \bibinfo{pages}{473}
  (\bibinfo{year}{2000}), \eprint{astro-ph/9911177}.

\bibitem[{\citenamefont{LoVerde et~al.}(2007)\citenamefont{LoVerde, Hui, and
  Gaztanaga}}]{LoVerde:2006cj}
\bibinfo{author}{\bibfnamefont{M.}~\bibnamefont{LoVerde}},
  \bibinfo{author}{\bibfnamefont{L.}~\bibnamefont{Hui}}, \bibnamefont{and}
  \bibinfo{author}{\bibfnamefont{E.}~\bibnamefont{Gaztanaga}},
  \bibinfo{journal}{Phys. Rev.} \textbf{\bibinfo{volume}{D75}},
  \bibinfo{pages}{043519} (\bibinfo{year}{2007}), \eprint{astro-ph/0611539}.

\bibitem[{\citenamefont{Christensen and Meyer}(2000)}]{Christensen:2000ji}
\bibinfo{author}{\bibfnamefont{N.}~\bibnamefont{Christensen}} \bibnamefont{and}
  \bibinfo{author}{\bibfnamefont{R.}~\bibnamefont{Meyer}}
  (\bibinfo{year}{2000}), \eprint{astro-ph/0006401}.

\bibitem[{\citenamefont{Christensen et~al.}(2001)\citenamefont{Christensen,
  Meyer, Knox, and Luey}}]{Christensen:2001gj}
\bibinfo{author}{\bibfnamefont{N.}~\bibnamefont{Christensen}},
  \bibinfo{author}{\bibfnamefont{R.}~\bibnamefont{Meyer}},
  \bibinfo{author}{\bibfnamefont{L.}~\bibnamefont{Knox}}, \bibnamefont{and}
  \bibinfo{author}{\bibfnamefont{B.}~\bibnamefont{Luey}},
  \bibinfo{journal}{Class. Quant. Grav.} \textbf{\bibinfo{volume}{18}},
  \bibinfo{pages}{2677} (\bibinfo{year}{2001}), \eprint{astro-ph/0103134}.

\bibitem[{\citenamefont{Knox et~al.}(2001)\citenamefont{Knox, Christensen, and
  Skordis}}]{Knox:2001fz}
\bibinfo{author}{\bibfnamefont{L.}~\bibnamefont{Knox}},
  \bibinfo{author}{\bibfnamefont{N.}~\bibnamefont{Christensen}},
  \bibnamefont{and} \bibinfo{author}{\bibfnamefont{C.}~\bibnamefont{Skordis}},
  \bibinfo{journal}{Astrophys. J.} \textbf{\bibinfo{volume}{563}},
  \bibinfo{pages}{L95} (\bibinfo{year}{2001}), \eprint{astro-ph/0109232}.

\bibitem[{\citenamefont{{Lewis} and {Bridle}}(2002)}]{LewBri02}
\bibinfo{author}{\bibfnamefont{A.}~\bibnamefont{{Lewis}}} \bibnamefont{and}
  \bibinfo{author}{\bibfnamefont{S.}~\bibnamefont{{Bridle}}},
  \bibinfo{journal}{\prd} \textbf{\bibinfo{volume}{66}},
  \bibinfo{pages}{103511} (\bibinfo{year}{2002}), \eprint{astro-ph/0205436}.

\bibitem[{\citenamefont{Kosowsky et~al.}(2002)\citenamefont{Kosowsky,
  Milosavljevic, and Jimenez}}]{Kosowsky:2002zt}
\bibinfo{author}{\bibfnamefont{A.}~\bibnamefont{Kosowsky}},
  \bibinfo{author}{\bibfnamefont{M.}~\bibnamefont{Milosavljevic}},
  \bibnamefont{and} \bibinfo{author}{\bibfnamefont{R.}~\bibnamefont{Jimenez}},
  \bibinfo{journal}{Phys. Rev.} \textbf{\bibinfo{volume}{D66}},
  \bibinfo{pages}{063007} (\bibinfo{year}{2002}), \eprint{astro-ph/0206014}.

\bibitem[{\citenamefont{Verde et~al.}(2003)}]{Verde:2003ey}
\bibinfo{author}{\bibfnamefont{L.}~\bibnamefont{Verde}} \bibnamefont{et~al.},
  \bibinfo{journal}{Astrophys. J. Suppl.} \textbf{\bibinfo{volume}{148}},
  \bibinfo{pages}{195} (\bibinfo{year}{2003}), \eprint{astro-ph/0302218}.

\bibitem[{\citenamefont{Gelman and Rubin}(1992)}]{gelman/rubin}
\bibinfo{author}{\bibfnamefont{A.}~\bibnamefont{Gelman}} \bibnamefont{and}
  \bibinfo{author}{\bibfnamefont{D.}~\bibnamefont{Rubin}},
  \bibinfo{journal}{Statistical Science} \textbf{\bibinfo{volume}{7}},
  \bibinfo{pages}{452} (\bibinfo{year}{1992}).

\bibitem[{\citenamefont{Gaztanaga et~al.}(2006)\citenamefont{Gaztanaga, Manera,
  and Multamaki}}]{Gaztanaga:2004sk}
\bibinfo{author}{\bibfnamefont{E.}~\bibnamefont{Gaztanaga}},
  \bibinfo{author}{\bibfnamefont{M.}~\bibnamefont{Manera}}, \bibnamefont{and}
  \bibinfo{author}{\bibfnamefont{T.}~\bibnamefont{Multamaki}},
  \bibinfo{journal}{Mon. Not. Roy. Astron. Soc.}
  \textbf{\bibinfo{volume}{365}}, \bibinfo{pages}{171} (\bibinfo{year}{2006}),
  \eprint{astro-ph/0407022}.

\end{thebibliography}

\vfill
\end{document}